\documentclass[aps,prd,amsmath,amssymb,reprint,superscriptaddress, nofootinbib,
]{revtex4-1}

\usepackage{hepunits}
\usepackage{graphicx}\usepackage{dcolumn}\usepackage{bm}\usepackage[mathlines]{lineno}\usepackage{lipsum}
\usepackage{xspace}
\usepackage[nolist]{acronym}
\usepackage{color}
\usepackage[caption=false]{subfig}
\usepackage{tikzscale}
\usepackage{circuitikz}
\usepackage{comment}

\graphicspath{{./},{images/}}

\usepackage{etoolbox}
\makeatletter
\patchcmd\linenumberpar{\@LN@parpgbrk}{\penalty\@LN@parpgpen\relax}{}{}
\makeatother

\begin{document}

\title[ABRACADABRA-10 cm]{Design and Implementation of the ABRACADABRA-10\,cm Axion Dark Matter Search}
\author{Jonathan~L.~Ouellet}
\email{ouelletj@mit.edu}
\affiliation{Laboratory for Nuclear Science, Massachusetts Institute of Technology, Cambridge, MA 02139, U.S.A.}

\author{Chiara~P.~Salemi}
\affiliation{Laboratory for Nuclear Science, Massachusetts Institute of Technology, Cambridge, MA 02139, U.S.A.}

\author{Joshua~W.~Foster}
\affiliation{Leinweber Center for Theoretical Physics, Department of Physics, University of Michigan, Ann Arbor, MI 48109, U.S.A.}

\author{Reyco~Henning}
\affiliation{University of North Carolina, Chapel Hill, NC 27599, U.S.A.}
\affiliation{Triangle Universities Nuclear Laboratory, Durham, NC 27708, U.S.A.}

\author{Zachary~Bogorad}
\affiliation{Laboratory for Nuclear Science, Massachusetts Institute of Technology, Cambridge, MA 02139, U.S.A.}

\author{Janet~M.~Conrad}
\affiliation{Laboratory for Nuclear Science, Massachusetts Institute of Technology, Cambridge, MA 02139, U.S.A.}

\author{Joseph~A.~Formaggio}
\affiliation{Laboratory for Nuclear Science, Massachusetts Institute of Technology, Cambridge, MA 02139, U.S.A.}

\author{Yonatan~Kahn}
\affiliation{Princeton University, Princeton, NJ 08544, U.S.A.}
\affiliation{Kavli Institute for Cosmological Physics, University of Chicago, Chicago, IL 60637, U.S.A.}

\author{Joe~Minervini}
\affiliation{Plasma Science and Fusion Center, Massachusetts Institute of Technology, Cambridge, MA 02139, U.S.A.}

\author{Alexey~Radovinsky}
\affiliation{Plasma Science and Fusion Center, Massachusetts Institute of Technology, Cambridge, MA 02139, U.S.A.}

\author{Nicholas~L.~Rodd}
\affiliation{Berkeley Center for Theoretical Physics, University of California, Berkeley, CA 94720, U.S.A.}
\affiliation{Theoretical Physics Group, Lawrence Berkeley National Laboratory, Berkeley, CA 94720, U.S.A.}

\author{Benjamin~R.~Safdi}
\affiliation{Leinweber Center for Theoretical Physics, Department of Physics, University of Michigan, Ann Arbor, MI 48109, U.S.A.}

\author{Jesse~Thaler}
\affiliation{Center for Theoretical Physics, Massachusetts Institute of Technology, Cambridge, MA 02139, U.S.A.}

\author{Daniel~Winklehner}
\affiliation{Laboratory for Nuclear Science, Massachusetts Institute of Technology, Cambridge, MA 02139, U.S.A.}

\author{Lindley~Winslow}
\email{lwinslow@mit.edu}
\affiliation{Laboratory for Nuclear Science, Massachusetts Institute of Technology, Cambridge, MA 02139, U.S.A.}
 \date{\today}

\begin{abstract}
The past few years have seen a renewed interest in the search for light particle dark matter. ABRACADABRA is a new experimental program to search for axion dark matter over a broad range of masses, \mbox{$10^{-12}\lesssim m_a\lesssim10^{-6}$\,eV}. ABRACADABRA-10\,cm is a small-scale prototype for a future detector that could be sensitive to QCD axion couplings. In this paper, we present the details of the design, construction, and data analysis for the first axion dark matter search with the ABRACADABRA-10\,cm detector. We include a detailed discussion of the statistical techniques used to extract the limit from the first result with an emphasis on creating a robust statistical footing for interpreting those limits.

\end{abstract}
 
\maketitle
\begin{acronym}
\acro{ABRA}[ABRACADABRA]{{\bf A} {\bf B}roadband/{\bf R}esonant {\bf A}pproach to {\bf C}osmic {\bf A}xion {\bf D}etection with an {\bf A}mplifying {\bf B}-field {\bf R}ing {\bf A}pparatus}
\acro{PSD}{power spectral density}
\acro{ALP}{Axion-like particles}
\acro{DM}{dark matter}
\acro{ADM}{Axion DM}
\acro{DFT}{discrete Fourier transform}
\acro{FLL}{flux-lock feedback loop}
\acro{SNR}{signal-to-noise ratio}
\acro{LEE}{look-elsewhere effect}
\acro{TS}{test-statistic}
\acro{POM}{polyoxymethylene}
\acro{PTFE}{polytetrafluoroethylene}
\acro{MC}{Monte Carlo}
\acro{WIMP}{Weakly Interacting Massive Particle}
\end{acronym}

\newcommand{\ABRA}{\ac{ABRA}\xspace}
\newcommand{\PSD}{\ac{PSD}\xspace}
\newcommand{\abra}{ABRACADABRA-10\,cm\xspace}
\newcommand{\ALP}{\ac{ALP}\xspace}
\newcommand{\ALPs}{\ac{ALP}s\xspace}
\newcommand{\DM}{\ac{DM}\xspace}
\newcommand{\ADM}{\ac{ADM}\xspace}
\newcommand{\gagg}{\ensuremath{g_{a\gamma\gamma}}\xspace}
\newcommand{\rhoDM}{\ensuremath{\rho_{\rm DM}}\xspace}
\newcommand{\DFT}{\ac{DFT}\xspace}
\newcommand{\FLL}{\ac{FLL}\xspace}
\newcommand{\SNR}{\ac{SNR}\xspace}
\newcommand{\LEE}{\ac{LEE}\xspace}
\newcommand{\TS}{\ac{TS}\xspace}
\newcommand{\POM}{\ac{POM}\xspace}
\newcommand{\PTFE}{\ac{PTFE}\xspace}
\newcommand{\MC}{\ac{MC}\xspace}
\newcommand{\WIMP}{\ac{WIMP}\xspace}

\newcommand{\Px}{\ensuremath{\bar{\mathcal{F}}}}
\newcommand{\Pten}{\ensuremath{\bar{\mathcal{F}}_\mathrm{10M}}\xspace}
\newcommand{\Pone}{\ensuremath{\bar{\mathcal{F}}_\mathrm{1M}}\xspace}
\newcommand{\Phun}{\ensuremath{\mathcal{F}_\mathrm{100k}}\xspace} 
\section{Introduction}
In recent years, the absence of a compelling direct detection of \DM in accelerator and \WIMP searches has reignited the search for \ADM. The coincidence of new developments in quantum sensors and quantum information technology has stoked this reawakened interest, and the past few years have seen a wealth of new experimental ideas and approaches that are beginning to revolutionize the field \cite{Marsh:2017hbv,Battaglieri:2017aum}. While most \ADM searches have traditionally focused on a narrow mass range from $10\lesssim m_a \lesssim 100\,\mu$eV, recent theoretical work has made a compelling case for \ADM in the mass range $m_a\lesssim 1\,\mu$eV \cite{Tegmark:2005dy,Davoudiasl:2015vba,Graham:2018jyp,Agrawal:2017eqm,2010PhRvD..81f3508V,Co:2016fln,DiLuzio:2018gqe,Arvanitaki:2009fg,Svrcek:2006yi,Agrawal:2017cmd,Farina:2016tgd}.

The \abra experiment has recently released results of the first direct detection search for \ADM below 1\,$\mu$eV \cite{ABRAFirstResults}. The design of the experiment was motivated by the proposal of \cite{ABRA2016}, and is based on measuring the coupling of \ADM to electromagnetism -- similar to experiments probing different mass regimes like the long-running ADMX \cite{Asztalos2001,ADMX2018} and HAYSTAC \cite{HAYSTAC2018a}. In this lower mass range, the axion field $a$, in the presence of a large magnetic field can be thought of as an induced effective current
\begin{equation}
\mathbf{J}_{\rm eff} = \gagg\frac{\partial a}{\partial t} \mathbf{B}\,,
\label{eqn:EffectiveCurrent}
\end{equation}
where \gagg is the axion-photon coupling. This current sources a small AC magnetic field that can be measured with a sensitive enough magnetometer. 

\abra is a prototype detector for a new search approach, and its implementation contains novel elements that have not previously been used in ultralight dark matter searches:
\begin{itemize}
    \item A toroidal magnet geometry, with the detection element placed in the near-zero-field region -- the first operational non-microwave cavity sub-eV \ADM search;
    \item A broadband readout mode involving continuous-stream data-taking for $\sim 10^{6}$ seconds (roughly 1 month), and several compression techniques to mitigate the total data storage requirements while preserving the desired signal bandwidth;
    \item A calibration technique where a signal is injected through current in a calibration loop in a similar geometry as the expected axion signal;
    \item A data analysis pipeline tailored to the expected statistics of the axion field in the quasistatic regime, where the signal is best described as a flux power spectral density rather than photon-counting with the added constraint that ``rescanning'' is prohibitively time-consuming.
\end{itemize}

In this paper, we provide context and additional details for each of these novel elements and their specific implementation in \abra. In Section~\ref{sec:DetectorDesign}, we describe the design and construction of the toroidal \abra detector. We describe the data collection approach used for the broadband readout in Sec~\ref{sec:DataCollection}, and describe the calibration of the detector in Sec.~\ref{sec:Calibration}. In Sec.~\ref{sec:DataAnalysis}, we describe the data analysis and limit extraction approach used for our broadband search. We conclude by commenting on the improvements and modifications necessary for scaling up the \abra prototype to an experiment capable of probing QCD axion couplings. 
\section{Detector Design and Construction}
\label{sec:DetectorDesign}
\begin{figure*}
\centering
\subfloat[\label{fig:ToroidWedges}]{\centering\includegraphics[width=.3\textwidth]{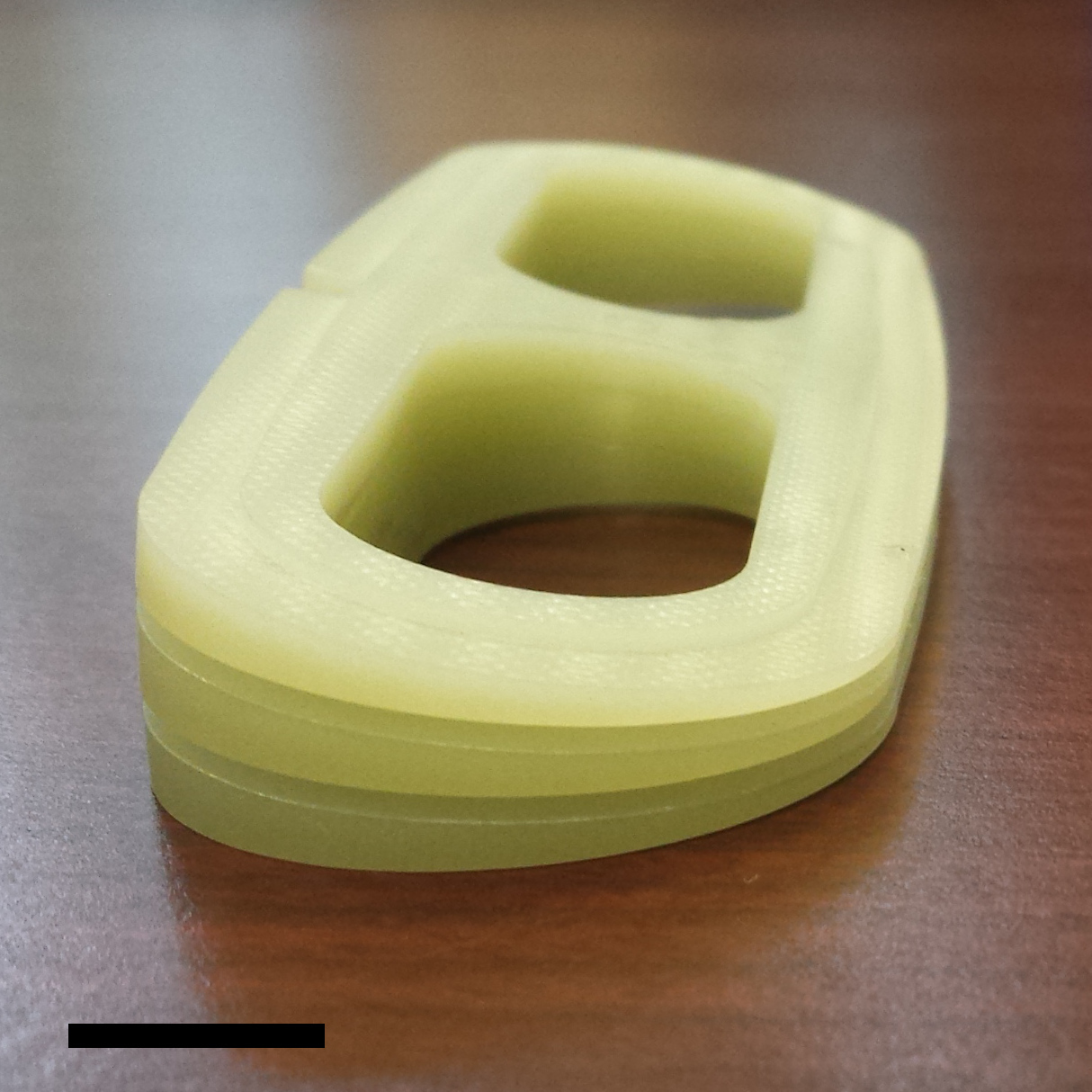}}\hspace{3cm}\subfloat[\label{fig:LoopsCutAway}]{\centering\includegraphics[width=.3\textwidth]{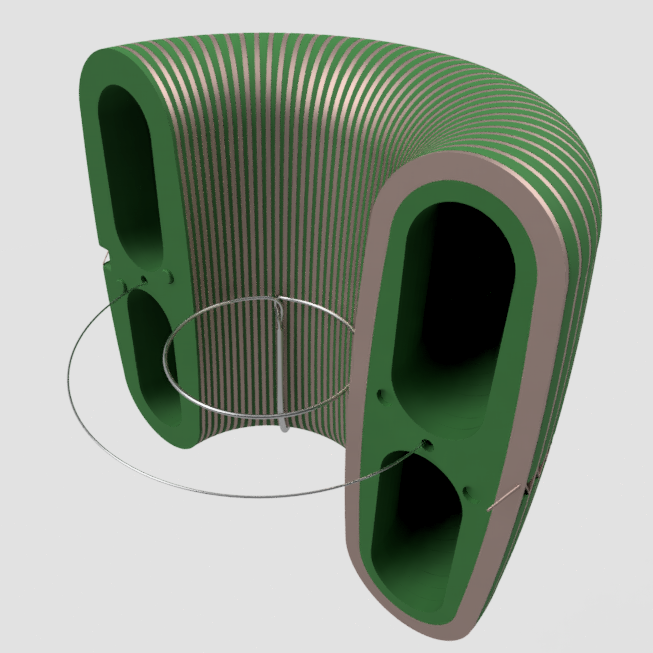}}\\
\subfloat[\label{fig:SupportStructure}]{\centering\includegraphics[width=.3\textwidth]{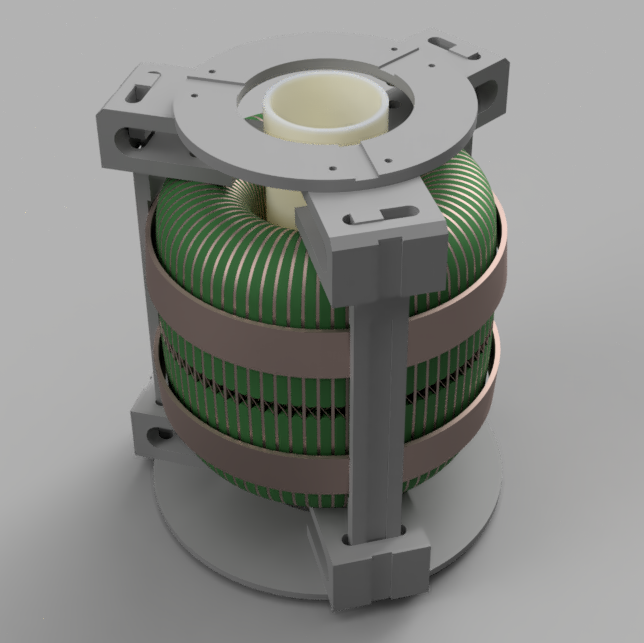}}\hspace{3cm}\subfloat[\label{fig:ABRAAssembledPhoto}]{\centering\includegraphics[width=.3\textwidth]{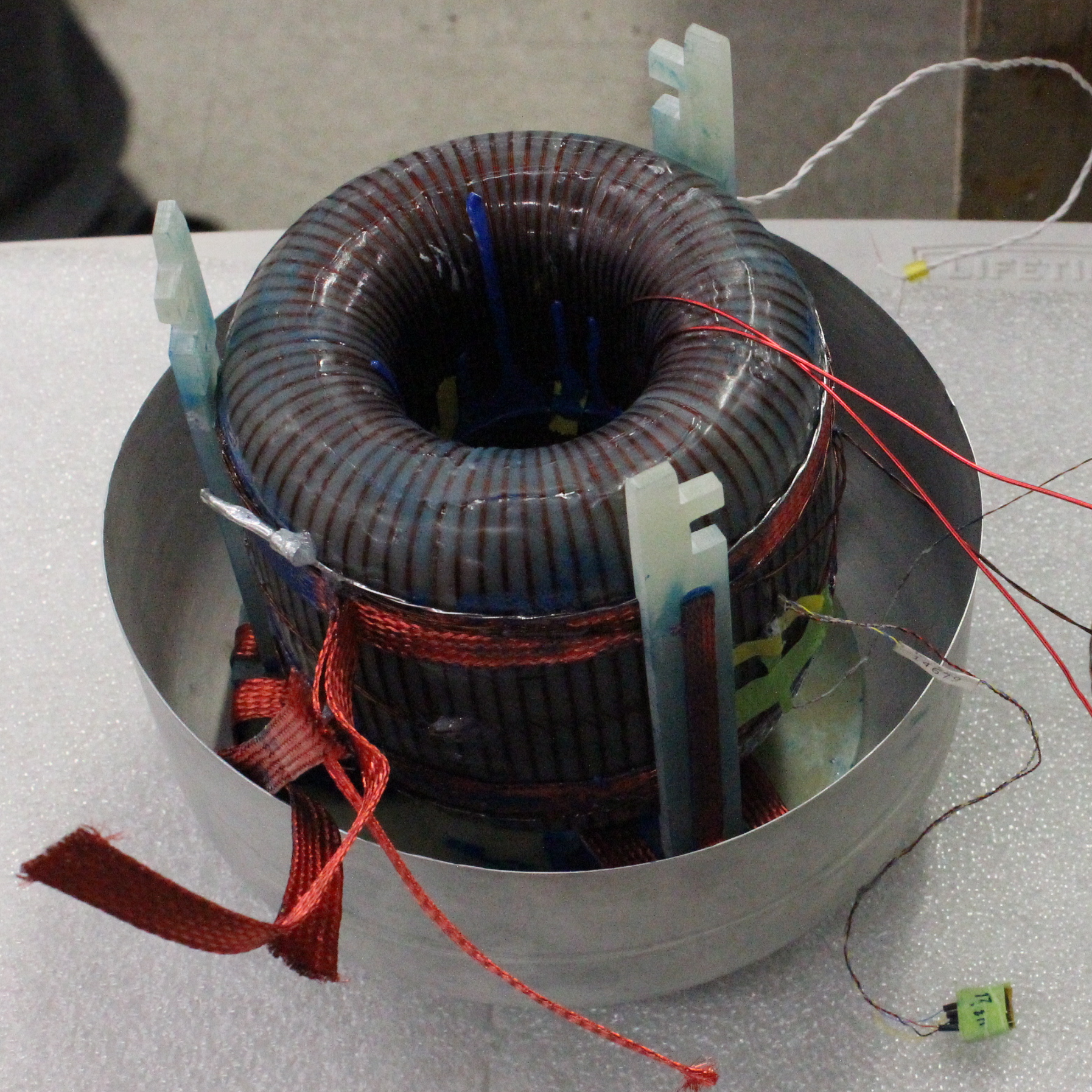}}\caption{(a) Three of the 80 Delrin wedges that form the toroid structure stacked together. The black bar indicates a $\approx 1$\,cm scale. (b) Cutaway rendering of the toroid with the 1\,mm diameter wire pickup loop in the center. A 0.5\,mm diameter wire runs through the center of field region to form the calibration loop. Toroid height is $\approx 12$\,cm. (c) Rendering of the \abra support structure. The pickup loop is supported by a PTFE (white) tube through the center. The magnet is supported by an outer G10 support structure and thermalized with two copper bands. (d) Photo of the assembled \abra, with the top of the superconducting shield and support structure removed.}
\label{fig:AssemblyPhotos}
\end{figure*}

The \abra detector and setup can be split into six separate systems: the toroidal magnet, the magnet support infrastructure and shielding, the pickup loop circuit, the SQUID electronics, the calibration circuit, and finally the cryostat and detector support infrastructure. In this section, we discuss the design and construction of each.

The expected \SNR of \abra can be written approximately as 
\begin{equation}
{\rm SNR} = \gagg\sqrt{\rhoDM}\mathcal{G}VB_{\rm max}\left(\frac{M_{\rm in}}{L_T}\right)\frac{\left(\tau t\right)^{\frac14}}{S_{\Phi\Phi}^{1/2}}\,,
\end{equation}
where $V$ is the volume of the toroid, $\mathcal{G}$ is a geometric factor, $B_{\rm max}$ is the max field inside the toroid, $M_{\rm in}$ is the inductive coupling of the SQUID, and $L_T$ is the total inductance of the readout circuit. 
Here, we assume that the integration time $t$, exceeds the axion coherence time, $\tau$. The final parameter of importance is the flux noise level, $S_{\Phi\Phi}^{1/2}$, typically measured in $\mu\Phi_0/\sqrt{\rm Hz}$. The relevant parameters are summarized in Table~\ref{tab:DetectorDesign}.

\subsection{The Toroidal Magnet}
\label{sec:TheMagnet}

The magnet structure is built around 80 identical Delrin wedges, (see Fig.~\ref{fig:ToroidWedges}). When glued together, they create a toroidal frame with an inner radius of 3\,cm and an outer radius of 6\,cm, with a total height of 12\,cm. The total volume of magnetic field is $V\approx890$\,cm$^3$.

The magnet current is carried by a NbTi(CuNi) wire which is wound 1,280 times around the magnet. Between each pair of wedges is a groove that has 16 winds of wire, laid down in pairs 8 layers deep. The wire is held in place with epoxy. To cancel the azimuthal current, the toroid is counterwound. 

The toroid was wound by Superconducting Systems, Inc \cite{SSI} in three separate pieces, with three separate lengths of NbTi(CuNi) wire. The pieces are then glued together and the wires are connected together with two superconducting crimps. These crimps are then attached to the outside of the toroid. These joints could create small stray fields which contribute to the backgrounds for the axion search, but unfortunately could not be avoided in the construction.

The toroid is mounted in a dilution refrigerator (described below) and cooled to $\lesssim1$\,K. The NbTi(CuNi) wire superconducts below $\lesssim9$\,K. We charge the magnet by injecting a 121\,A current into the toroid. Once charged, we turn off heat to a superconducting switch (located away from the magnet) which then locks the current into the magnet. The current source is disconnected from the charging leads on top of the refrigerator.

When fully charged, the maximum field in the magnet is $B_{\rm max}=1$\,T. This was confirmed with a Hall probe to a precision of $\sim1$\%, with the uncertainty coming from uncertainty in position of the probe in the field. Once in persistent mode, we observed no decay in the field to a precision of $\lesssim0.1$\% on the scale of 1 week. The Hall sensor was removed before normal data taking.

\subsection{Support Infrastructure}
\label{sec:ToroidSupport}

The toroid is mechanically supported by a G10 frame held together with nylon bolts (see Fig.~\ref{fig:SupportStructure}). The goal of this structure was to rigidly mount the toroid in place, while minimizing the amount of non-superconducting metal near the magnet. The one exception to this is the copper straps which wrap around the outside of the toroid that provide the required thermalization to cool the magnet. These straps undoubtedly contribute some level of noise for our axion search, though in the current setup it is not the dominant noise source. In the future we will search for alternative thermalization approaches.

The entire toroid and support structure are mounted inside the external shield (see Fig.~\ref{fig:ABRAAssembledPhoto}). The shield consists of a spun copper can that has been coated inside and out with a 25-75\,$\mu$m layer of tin, for a total thickness of $\approx$1\,mm. The copper provides good thermal conductivity to minimize thermal gradients across the shield. It also provides the thermal conductivity to the copper straps which cool the magnet. Once below 3.7\,K, the tin becomes superconducting and expels environmental magnetic fields and acts as a shield against electromagnetic interference. Optimizing and characterizing this external shielding will be the subject of future work.

The external shield is built in two hemispheres (top and bottom) which have $\approx12$\,mm of vertical overlap when assembled. There is a small gap in one location between the inner and outer shield through which the magnet wires, pickup loop wires and calibration loop wires pass as three sets of twisted pairs. The shield halves are connected with a layer of solder and epoxy to ensure both electrical and mechanical connection.

A 12\,mm thick aluminum top plate is epoxied to the top of the top shield and acts as the contact point for the thermalization to the rest of the cryostat and mechanical mounting point to the vibration isolation system (see below). The aluminum plate is electrically isolated from the shield to minimize grounding loops. A thermometer is epoxied to the outside of the bottom shield which monitors the temperature of the farthest point from the thermalization. However, during data taking this thermometer is not active.

\begin{table}
\centering
\caption{Summary of the \abra detector design parameters.}
\label{tab:DetectorDesign}
\begin{tabular}{lcc}
\hline
\hline
Pickup Loop Radius & $R_p$ & 20.1\,mm \\ Pickup Loop Wire Diameter & $r_p$ & 1.0\,mm \\
\hline
Magnet Inner Radius & $R_{\rm in}$ & 30\,mm \\
Magnet Outer Radius & $R_{\rm out}$ & 60\,mm \\
Magnet Height & $h$ & 120\,mm \\
\hline
Magnet Max Field & $B_{\rm max}$ & 1.0\,T\\
Geometric Factor & $\mathcal{G}_V$ & 0.027\\
\hline 
Pickup Loop Inductance & $L_p$ & 95.5\,nH \\
SQUID Input Inductance & $L_{\rm in}$ & 150\,nH\\
SQUID Inductive Coupling & $M_{\rm in}$ & 2.5\,nH\\
\hline
\hline
\end{tabular}
\end{table}

\subsection{Pickup Loop Circuit}
\label{sec:PickupLoop}
\begin{figure}
\centering
\includegraphics{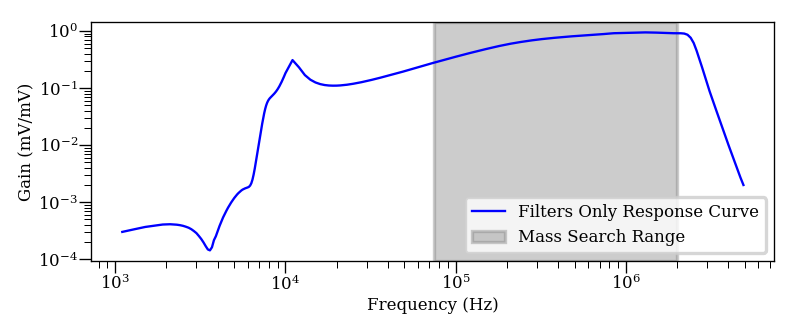}
\caption{Gain of the combined high-pass and anti-aliasing filters. All spectra are corrected for this response function -- unless otherwise noted. Measured in-situ, using injected signals at different frequencies. This also defines the usable range of data. For our search we use the range 75\,kHz -- 2\,MHz.}
\label{fig:FiltersResponse}
\end{figure}

The pickup loop measures the magnetic field in the center of the toroid -- a region that should have zero field in the absence of an axion signal. The time-averaged magnitude of the flux through the pickup loop due to $\mathbf{J}_{\rm eff}$ can be written as
\begin{equation}
|\Phi_a|^2 = \gagg^2\rhoDM V^2\mathcal{G}_V^2B_{\rm max}^2\equiv A\,,
\label{eq:pickup_flux_power}
\end{equation}
where \rhoDM is the \DM density, $V$ is the volume of the magnet, $B_{\rm max}$ is the maximum field in the magnet and $\mathcal{G}$ is a geometric factor. The pickup loop itself consists of a 1\,mm diameter solid NbTi wire wrapped around the outside of a 5.5\,cm diameter \PTFE tube that is 18.1\,cm tall. 
The geometric factor $\mathcal{G}_V$ weights the effective current in Eqn.~\ref{eqn:EffectiveCurrent}, by the contribution to the flux through the pickup loop. This can be written as
\begin{equation}
\small
\mathcal{G}_V \equiv \frac{1}{B_{\rm max}V}\left|\int_{\rm Loop} dA \, \int_{\rm Toroid} dV' \ \!\frac{\mathbf{B}(\mathbf{r}')\times(\mathbf{r}'-\mathbf{r})}{|\mathbf{r}'-\mathbf{r}|^3}\cdot\mathbf{\hat{n}}\right|
\label{eq:GVdef}
\end{equation}
where $\mathbf{\hat{n}}$ is the normal to the plane of the pickup loop, and the integrals are taken over the area enclosed by the pickup loop and the volume of the toroid. The integrand is reminiscent of the Biot-Savart law, with the current taken to be the axion-induced effective current $\mathbf{J}_{\rm eff}$ which follows lines of $\mathbf{B}$ \cite{ABRA2016}.\footnote{In the notation of \cite{ABRA2016}, $\mathcal{G}_V = V_B/V$.} For the \abra geometry, we calculate this using a COMSOL \cite{COMSOL} simulation to be $\mathcal{G}_V=0.027$. 
The two wire leads from the pickup loop, which consists of the same wire as the loop, are twisted into a twisted pair configuration and run out under the bottom of the toroid through the gap in the shield. Once outside of the shield, the wires run $\approx15$\,cm inside a stainless steel mesh sleeve. At this point, the 1\,mm wires are joined to 75\,$\mu$m twisted-pair PFA-insulated wire with superconducting crimped solder. The 75\,$\mu$m wires run for $\approx$1\,m inside hollow superconducting solder capillaries \cite{Solder-Paper} to the input of the SQUIDs mounted on the 700\,mK (Still) stage of the cryostat. The SQUIDs have an input inductance of 150\,nH to match the calculated inductance of the pickup loop of $L_p=95.5$\,nH. Including the inductance of the wires, the total design inductance of the pickup loop circuit is $\approx$550\,nH. However, measuring the inductance of the circuit yielded a value closer to $L_T\approx3.3\,\mu$H, we discuss this further in Sec.~\ref{sec:Calibration}. The data presented in \cite{ABRAFirstResults} was taken in a broadband readout configuration with no resonant amplifier.

\subsection{SQUID Setup And Readout}
\label{sec:SQUIDReadout}

The first stage was readout with a Magnicon two stage SQUID current sensor \cite{Magnicon,Drung2007}. The SQUID is operated at a temperature of 870\,mK and has typical flux noise floor of $0.6\,\mu\Phi_0^2/\mathrm{Hz}$. The inductive coupling between the input coil of the SQUID and the SQUID is $M_{\rm in} = 2.52$\,nH. The SQUID is operated with the Magnicon XXF-1 electronics in \FLL mode with a SQUID flux to voltage conversion of $\partial V/\partial \Phi_S = 1.29$\,V/$\Phi_0$. In \FLL mode, the response of the SQUID is linear over the dynamic range of the amplifier $\pm11$\,V or $\approx\pm8.5\,\Phi_0$, however, this comes at the cost of limiting the bandwidth of the system to $\approx6$\,MHz. 

The output voltage of the SQUID was recorded with an AlazarTech 9870 8-bit digitizer \cite{AlazarTech}. To achieve the needed voltage precision we use the smallest available input range of $\pm40$\,mV which leads to a typical digitizer noise floor of $3.5\times10^{-9}$\,mV$^2$/Hz. However, due to the large background below $\sim20$\,kHz (see Fig.~\ref{fig:SQUIDAccelCorrelation}), we must first pass the signal through a 10\,kHz high-pass filter to prevent railing the digitizer. Additionally, we use a 2.5\,MHz anti-aliasing filter. The frequency response of these two filters is shown in Fig.~\ref{fig:FiltersResponse}. These filters define the usable range of data for our axion search of 50\,kHz -- 3\,MHz. It is worth pointing out that even though the gain is less than unity over the majority of our search range, both the signal and dominant noise is scaled by this gain, so the \SNR is unchanged. 

The digitizer is clocked to a Stanford Research System FS725 Rb frequency standard, with a ten-second Allan variance of $<10^{-11}$.

\subsection{Calibration Circuit}
\label{sec:CalibrationLoop}

\begin{figure}
\includegraphics[width=.5\textwidth]{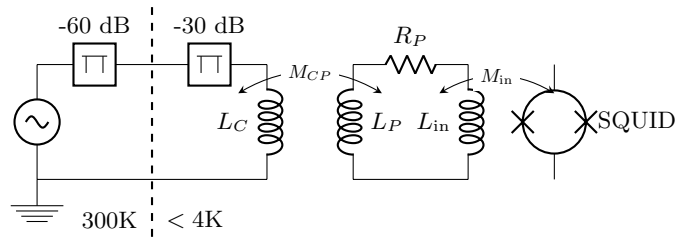}
\caption{A conceptual diagram of the \abra calibration circuit. The calibration loop, $L_C\approx300$\,nH, is concentric with the pickup loop, $L_P=95.5$\,nH. The circuit is plugged into the SQUID with input inductance $L_{\rm in}\approx150$\,nH. The parasitic resistance in the circuit is measured as $R_P\approx13\,\mu\Omega$.}
\label{fig:CalibrationCircuit}
\end{figure}

We measure the end-to-end gain using a calibration system. It consists of the 0.5\,mm diameter NbTi wire passing through the body of the toroid (i.e. in the magnetic field region), creating a 9\,cm diameter loop concentric with and in the same plane as the pickup loop (see Fig.~\ref{fig:LoopsCutAway}). This wire runs out of the detector shield as a twisted pair and then into an RG196 coaxial cable. This cable is connected to a 30\,dB attenuator at the 4\,K stage and then continues up to the top of the cryostat and through a BNC feed-through out of the vacuum region. During data taking, this BNC is left unplugged, and the attenuator contributes noise from a 50\,$\Omega$ resistor at 4\,K, which is well below our current noise level. We calculate the mutual inductance between the calibration and pickup loops both with an analytic calculation based on the geometry, as well as with a COMSOL \cite{COMSOL} simulation. These values agree and predict a mutual inductance of $M_{CP}=19.3$\,nH. 

While calibrating, we typically add an additional 60\,dB of warm attenuation for a total of 90\,dB of attenuation to get the signal to reasonable size. All attenuators are impedance matched at 50\,$\Omega$. However, the output of the final attenuator is shorted by the calibration loop, which has an inductance of $\approx$300\,nH; for frequencies below $\approx30$\,MHz this causes it to behave as a current source driving a current through the calibration loop with amplitude independent of frequency. A wiring schematic of the calibration circuit is shown in Fig.~\ref{fig:CalibrationCircuit}.

\subsection{Cryostat and Detector Suspension}
\label{sec:Cryostat}

\begin{figure}
\centering
\includegraphics{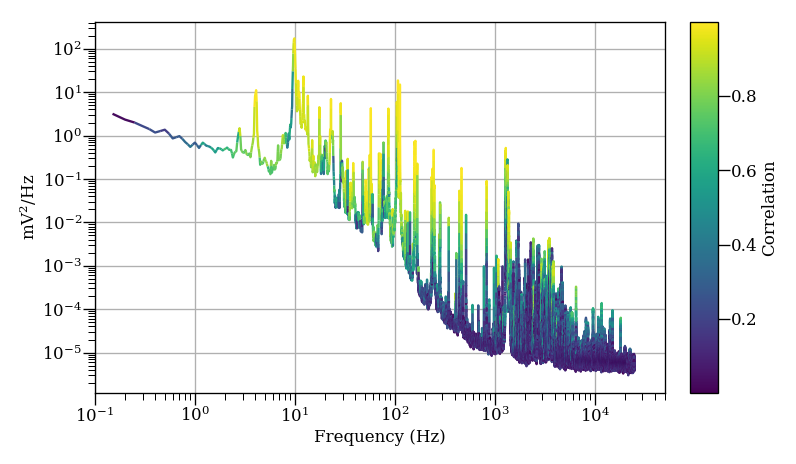}
\caption{Low frequency SQUID spectra from \abra taken with an accelerometer attached the 300\,K plate. The spectrum is that of the SQUID output, with the degree of correlation with the accelerometer indicated by color (i.e. the correlation coefficient). The accelerometer begins to lose sensitivity above a few kHz, so it is not clear from this measurement how far up the correlation continues. These data were taken with a larger dynamic range on the digitizer, so have a relatively high ADC noise floor of $\sim5\times10^{-6}$\,mV$^2$/Hz. (Data taken without signal shaping filters.)}
\label{fig:SQUIDAccelCorrelation}
\end{figure}

The \abra detector is mounted inside an Oxford Instruments Triton 400 dilution refrigerator. It is mechanically supported by the detector suspension system. This consists of a 0.038'' Kevlar thread which attaches to a vented bolt screwed into the center hole of the top aluminum mounting plate on \abra. The thread runs $\sim1.5$\,m up through the various cold stages of the cryostat to a steel spring which supports the weight of the detector. The steel spring has a spring constant of $k\approx20.4$\,N/m and connects to a hook which is mounted about 1\,m above the 300\,K plate of the cryostat. The hook is at the top of a 1\,m long vacuum tube which is rigidly connected to the 300\,K plate. Due to the poor thermal conductivity of Kevlar, the spring and top end of the thread stay at 300\,K while the bottom of the thread is cooled to $\lesssim$1\,K with the detector.

The detector is thermalized to the coldest stage of the cryostat through four 10\,mm wide 75\,$\mu$m thick copper ribbons. Specifically, they are attached to the Mixing Chamber plate of the cryostat and then to the aluminum top plate of the detector. They are mounted with significant slack to minimize vibration through these ribbons. They are electrically isolated from the top of the detector by using thin Kapton pads between the copper ribbon and the aluminum plate and connected with Nylon bolts.  

The detector suspension system is designed to act like a pendulum which rolls off lateral vibration above frequencies of $f\approx0.4$\,Hz and in the vertical direction above frequencies of $f\approx0.3$\,Hz. The operating frequency of the pulse tube is 1.4\,Hz and creates one of the main vibrational noise sources in the \abra data. It is clear from Fig.~\ref{fig:SQUIDAccelCorrelation}, that even with this suspension system, vibrational noise still plays a significant role in the \abra backgrounds, and is a future path for potential improvement.

To improve the magnetic shielding of the detector, we wrapped the cryostat in MuMetal shielding. As MuMetal performs best at room temperature, we only wrapped the outermost vacuum vessel. The vertical walls of the vessel were lined both inside and out with a $200\,\mu$m thick layer. The bottom of the vessel was covered with a single layer on the inside. The top of the vacuum vessel and cryostat were not covered with MuMetal due to all the instrumentation and cryostat infrastructure. We measured the DC magnetic field attenuation ex-situ to be a factor of $\sim 5-10$.

\section{Data Collection Procedure}
\label{sec:DataCollection}
\subsection{Axion Search Data}
\label{sec:AxionSearchData}

In \cite{ABRA2016}, the original proposal for a broadband search involved collecting time series data at a high sampling frequency continuously for months to years. However, this runs into practical disk space limitations. For example, one month sampled at 10\,MS/s would fill $\approx$26\,TB of disk space. This is maneagable, but would not scale well to a 1\,GS/s sampling rate for a full year. However, this sort of sampling is not necessary for resolving \ADM signals, where the expected signal width is given by $\Delta f/f\sim10^{-6}$. Instead, we take an approach that maintains the required spectral resolution, while minimizing the required disk space. 

For \abra, we sample continuously at 10\,MS/s. Once samples are pulled from the digitizer, the data follows two processing paths: transforming and downsampling. First, the samples are accumulated into a 10\,s buffer (of $10^8$ samples), which is then transformed via \DFT \cite{FFTW05} into a \PSD. Once the next 10\,s is available, it is transformed and its \PSD is then averaged with the first, and so on. This builds up an averaged \PSD, called \Pten, which has Nyquist frequency of 5\,MHz and frequency resolution of 100\,mHz. This spectrum would be able to resolve axion signals down to $m_a\sim 100$\,kHz with at least one bin width. After 80 averages (or 800\,s), the average spectrum is written to disk and the averaging is reset. This level of averaging was chosen as a balance between storage space and being able to resolve time variation of background noise.

In parallel with this, the 10\,MS/s time series is decimated by a factor of 10, to a 1\,MS/s time series. This data is then accumulated into a 100\,s buffer -- again of $10^8$ samples -- then transformed with a \DFT and converted into a \PSD. In this way, we build up a second averaged \PSD called, \Pone, with a Nyquist frequency of 500\,kHz and a frequency resolution of 10\,mHz. This spectrum would be able to resolve signals down to $m_a\sim10$\,kHz with at least one bin width. After 16 averages (or 1600\,s), the average spectrum is written to disk and the averaging is reset.

The data are then decimated by another factor of 10 and written directly to disk at a sampling rate of 100\,kS/s. Offline, we take the time series data and transform it as one $2.45\times10^{11}$ sample long \DFT to form a final spectrum, \Phun. Unlike the other spectra, \Phun is not averaged over multiple integration periods, but is instead a single \PSD with Nyquist frequency 50\,kHz and frequency resolution of $\approx408$\,nHz. This spectrum could be used for searches for axion signals down to below 1\,Hz, however, it is not used in the present analysis.

Each decimation step is done by first applying a top-hat filter with a 10-bin width, and then down-sampling by keeping every 10$^{\rm th}$ filtered sample. This approach was chosen because it is fast computationally, though it is not quite optimal. 
We collected data from July 16, 2018 through August 14, 2018, accumulating a total exposure of $T=2.45\times10^6$\,s or $24.5\times10^{12}$ samples. In total, the data consist of 3065 independent \Pten spectra and 1532 \Pone spectra as well as a 2.45$\times10^{11}$ continuous samples of 100\,kS/s data.
The total data footprint was about 3.8\,TB for an average write rate of 12.4 Mbps -- both of which are easily handled by a desktop PC.

\subsection{Magnet Off and Digitizer Noise Data}
\label{sec:MagnetOff}

We also perform a Magnet Off measurement to understand backgrounds that are not correlated with the magnet. This data was collected with the exact same procedure and hardware configuration as the Magnet On data. Neither the cryostat, nor the SQUIDs were stopped in between measurements. We started collecting Magnet Off data within a few days of stopping the Magnet On run. We collected Magnet Off data from August 18, 2018 through August 27, 2018, for a total of $8.00\times10^5$\,s of data.

We also collected $\approx16$\,h of digitizer noise data, with nothing plugged in to measure the noise level inherent to the ADC and computer.

\subsection{Data Quality}
\label{sec:DataQuality}

\begin{figure}
\centering
\includegraphics{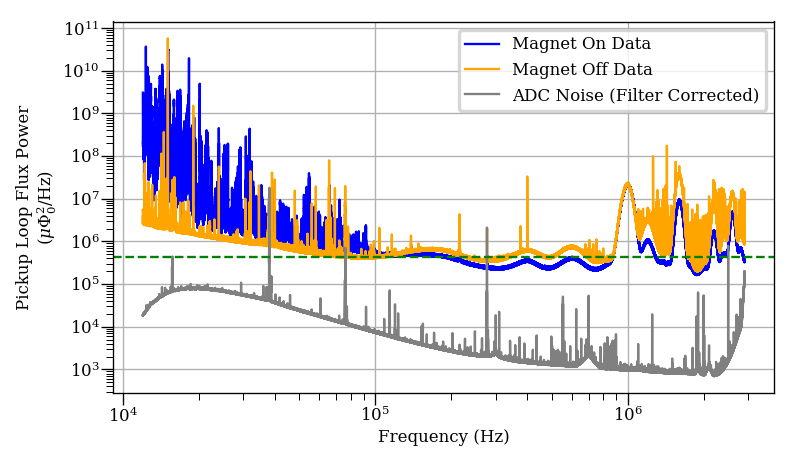}
\caption{Example \Pten SQUID spectra with magnet on (\emph{blue}) and off (\emph{orange}), along with the digitizer noise floor (\emph{gray}). SQUID spectra are averaged over $\approx$9\,h, digitizer data averaged over $\approx$16\,h. The typical SQUID noise floor is shown in green dashed line. Note: The spectra were collected at different times and some of the transient noise peaks are not seen in all spectra.}
\label{fig:ExampleSpectrum}
\end{figure}

Figure~\ref{fig:ExampleSpectrum} shows examples of spectra for magnet on and magnet off data. There are a few features of these spectra worth discussing. The region below 100\,kHz shows large noise spikes and a baseline increasing towards lower frequency. These spikes are generally too broad to be identified as \ADM, but instead are incoherent noise backgrounds. We also observe that this noise is significantly reduced when the magnet is turned off. We interpret this noise as due to vibration of the detector. In particular, this appears to be the high frequency tail of the noise in Fig.~\ref{fig:SQUIDAccelCorrelation}. The fact that it is reduced when the magnet is turned off implies that stray fields from the toroidal magnet are being seen by the pickup loop. We see that for $f>100$\,kHz this noise becomes sub-dominant; however, it is likely that it continues to higher frequency. This will present a major challenge for future detector configurations, including those with resonator readouts, that hope to lower the noise floor by many orders of magnitude. 

In the region from $100\,\mathrm{kHz}\lesssim f\lesssim850\,\mathrm{kHz}$, the noise is mostly flat with a few small broad bumps and is approximately consistent with the expected SQUID flux noise floor. We see a slow variation in this noise level over the month of data taking, associated with variations in the noise level of the SQUIDs. 

The region above $\sim850$\,kHz shows two effects: broad bumps with $\sim100$kHz widths and a forest of very narrow transient peaks. The broad peaks are due to an unknown and incoherent noise source that decreases our sensitivity in that frequency range. The origin of this noise will be the subject of future investigation, but for now we tolerate the decreased sensitivity. The forest of narrow transient peaks, on the other hand, present a larger problem. These peaks are $\lesssim$100\,mHz wide and actually narrower than we expect for an \ADM signal in this range. They are transient in time and appear to be correlated with working in the lab and with working hours. They were present for a portion of the time that we collected Magnet On data and all the time that we collected Magnet Off data. The transient nature and narrow width of these peaks imply that their source is likely from digital electronics turning on and off somewhere in, or near, the lab.

This transient noise was observed to be either present as a forest of many lines or completely absent. For our \ADM search, the easiest approach was to use this fact to tag and eliminate the effected periods of time. This could be done reliably by eye, but we use a more quantitative approach described in Sec.~\ref{sec:QualityCuts}. Though the lines only appeared at frequencies above $\sim850$\,kHz, we excluded all data from the tagged time intervals. In total, these cuts eliminated $\sim$30\% of the exposure.

\section{Calibration}
\label{sec:Calibration}

We quantify our detector response to a potential \ADM signal, by performing a set of calibration measurements. Each measurement involves injecting a series of AC signals with known amplitude and frequency into the calibration system described in Sec.~\ref{sec:CalibrationLoop}. We compare the power measured by our readout circuit to the power expected from the flux through the pickup loop generated by the calibration loop.

\begin{figure*}
\centering
\subfloat[\label{fig:CalibrationPeak}]{\centering\includegraphics{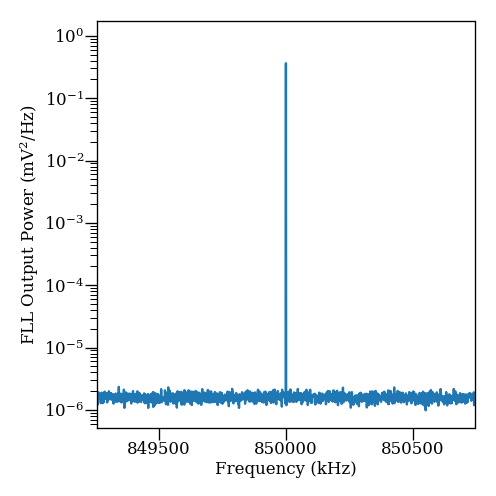}}\hspace{.5cm}
\subfloat[\label{fig:CalibrationScan}]{\centering\includegraphics{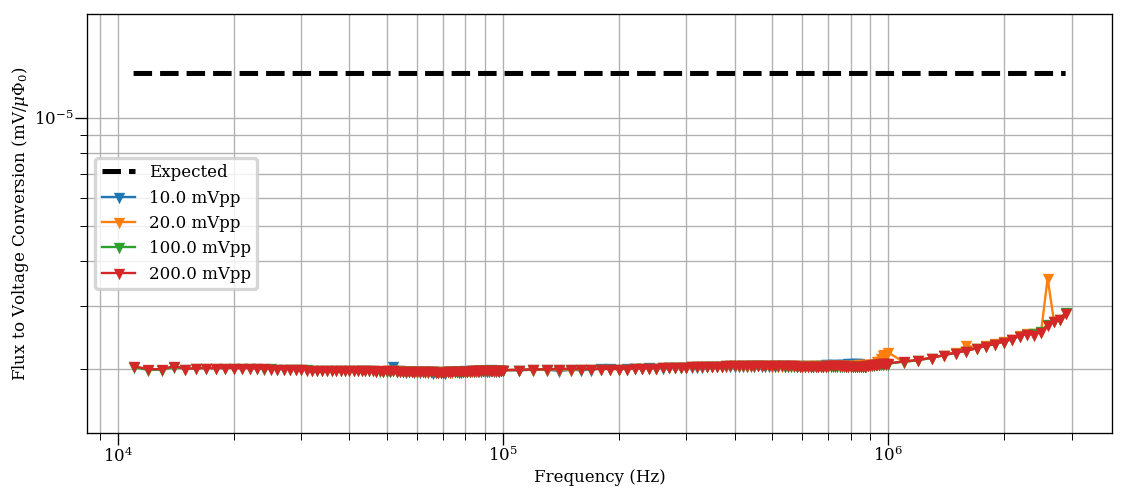}}\caption{(a) Example calibration peak at 850\,kHz with 10\,mVpp excitation and 90\,dB of attenuation. Bin width is 1\,Hz wide and all power is contained within a single bin. Output voltage is measured at the output of the amplifier electronics. (b) Measured detector response for four different input amplitudes taken with the magnet on. The measured gain is a factor of $\approx6.5$ below the expected response (dashed line at the top). The outlier in the 20\,mVpp spectrum is the result of a background line contributing power to the measured peak.}
\label{fig:CalibrationPlots}
\end{figure*}

The input signal is generated by a Stanford Research Systems SG380 signal generator, locked to the same Rb frequency standard as the digitizer. The SG380 has very low phase noise and is able to output a tone with very long coherence time (longer than our measurement time), such that the resulting peak in the \PSD was less than one frequency bin wide. For each amplitude and frequency, we perform a similar data collection to our axion search. We collect, transform and average 1\,s buffers to form an averaged \PSD. A zoomed example calibration line is shown in Fig.~\ref{fig:CalibrationPeak}. The resulting peaks are typically only one bin wide. We measure the power in each calibration peak and compare this to the expected flux power generated by the calibration loop. 

We perform this procedure for between 120 and 200 frequencies from 10\,kHz to 3\,MHz, and for four different input amplitudes: 10\,mVpp, 20\,mVpp, 100\,mVpp and 200\,mVpp. The resulting gain spectrum is shown in Fig.~\ref{fig:CalibrationScan}, and shows good agreement between the different input amplitudes.  We perform the calibration measurement before the \ADM search run, between the Magnet On run and the Magnet Off run, and again after the Magnet Off run. The resulting calibration curves were very consistent in time and did not depend on whether the magnet was on or off. For the present search, we determine our final calibration by interpolating the 200\,mVpp data taken with the magnet on (red curve in Fig.~\ref{fig:CalibrationScan}).

From Fig.~\ref{fig:CalibrationScan}, we see that our measured gain is a factor of $\approx6.5$ below what is expected based on the calculated circuit inductance. We tested each element of the calibration circuit and determined that the discrepancy came from the pickup loop side of the measurement. We determined the factor of $\approx$6.5 to most likely come from parasitic inductance in the readout circuit. This degrades the overall sensitivity of our axion search, and is the focus of future upgrades. 
\section{Axion Search and Limit Extraction}
\label{sec:DataAnalysis}

For the present analysis, we restrict our axion search to the frequency range $75\,\mathrm{kHz} < f < 2\,\mathrm{MHz}$ or axion mass range $0.31 <m_a<8.3$\,neV. We therefore do not include the \Phun spectrum in the rest of this analysis, as it has a Nyquist frequency of 50\,kHz. We use the \Pten spectra to search the range from $500\,\mathrm{kHz} < f < 2\,\mathrm{MHz}$, and the \Pone spectra to search the range $75\,\mathrm{kHz}<f<500\,\mathrm{kHz}$. In this way, a potential signal would be  covered by at least 10 frequency bins at all frequencies. We further average the averaged spectra \Pten and \Pone to contain 3200 and 480 averages respectively. This decreases our ability to resolve time variations in our background noise to $\approx$9\,h and $\approx$18\,h, respectively. This step is not necessary for our analysis and is purely to decrease the computational resources required by a factor of $\approx40$. After this, our 1 month of data collection is spanned by 75 \Pten spectra and 37 \Pone spectra.\footnote{We recycle the notation because we have only changed the number of spectra contributing to the average, but otherwise, they are conceptually equivalent.}

Our data analysis procedure closely follows the method introduced in \cite{Foster2018}.  Our expected signal is a narrow peak in the pickup loop \PSD above the noise background, with a width $\Delta f/f\sim10^{-6}$ arising from the \ADM velocity dispersion. The challenge in a broadband search such as this is that we are scanning a large number of mass points without the benefit of being able to efficiently `rescan' mass points with possible signal detections. As such, we need to be thorough with our statistical modeling, as at least some points are likely to populate the tails of any distribution. In this section, we describe the statistical modeling of our expected signal and background, as well as a data quality cut for tagging the periods of time when transient noise causes the data to look neither like signal nor background.

\subsection{Likelihood Analysis}
\label{sec:Likelihood}

The local ADM field can be thought of as arising from a partially coherent sum over a very large number of individual axion particles, where the phases of each particle are randomly distributed. As a result, the expected signal power in each frequency bin is drawn from an exponential distribution. When averaged over $N_{\rm avg}$ independent PSDs, the signal in each frequency bin $k$ will follow an Erlang distribution. When combined with background noise that is incoherent and Gaussian distributed in the time domain, the resulting \PSD data is still Erlang-distributed \cite{Foster2018}. Accordingly, for a single averaged \PSD, our combined signal-plus-background model prediction in each bin is an Erlang distribution,
\begin{equation}
 P(\Px_k; N_{\rm avg}, \lambda_k) 
 = \frac{N_{\rm avg}^{N_{\rm avg}}}{(N_{\rm avg}-1)!} \frac{(\Px_k)^{N_{\rm avg}-1}}{\lambda_k^{N_{\rm avg}}} 
 e^{-\frac{N_{\rm avg} \Px_k}{\lambda_k}}\,,
\end{equation}
with shape parameter $N_{\rm avg}$ and mean $\lambda_k = s_k + b$, where
\begin{eqnarray}
s_k = 
\left\{ \begin{array}{lc}
A \left. \frac{\pi f(v)}{m_a v} \right|_{v = \sqrt{4\pi f_k/m_a - 2}} & f_k > m_a/2\pi\,, \\
0 & f_k \leq m_a/2\pi\,,
\end{array} \right.
\label{eq:skdef}
\end{eqnarray}
and $b$ is the expected background power. Here, $A$ denotes the combination of parameters that control the signal strength, defined in Eqn.~\ref{eq:pickup_flux_power}. We assume $f(v)$ is given by the Standard Halo Model (SHM), with velocity dispersion $v_0 = 220$\,km/s, and $v_{\rm obs} = 232$\,km/s the DM velocity in the Earth frame \cite{McMillan:2009yr} and $\rhoDM=0.4$\,GeV/cm$^3$ \cite{Catena:2009mf,Iocco:2011jz}.

We build an analysis over a set of $\mathcal{N}$ averaged spectra $\Px_j$, each one an average over $N_{\rm avg}$ individual \PSD. For example, the analysis of the \Pten spectra, we have $\mathcal{N}=75$ averaged spectra, where each averaged spectrum $\Px_j$ is an average over $N_{\rm avg}=3200$ {\PSD}s (with the possible exception of the final averaged spectrum which usually has fewer {\PSD}s contributing). We search for an axion signal at mass $m_a=f/(2\pi)$, by restricting our search to a window containing frequency bins from $f_{k_i(m_a)}=m_a/(2\pi)$ to $f_{k_f(m_a)}=(1 + 4v_0^2)f_{k_i(m_a)}$ -- approximately 8 times the width of the expected signal. Since incoherent background noise varies on frequency scales much larger than this, we can approximate the background noise level in this window as independent of frequency. We tested that our final results were insensitive to the precise choice of this window width. We account for long term variability in our noise levels by allowing the expected background level to vary from one averaged spectrum $\Px_j$ to the next; we denote $\mathbf{b}=\{b_1,b_2,\dots,b_\mathcal{N}\}$ to be these background values, which we treat as nuisance parameters. The expected axion signal strength $A$ is constant across our data taking period and thus is the same for each $\Px_j$. For a given axion mass point $m_a$, signal strength $A$ and background values $\mathbf{b}$, we calculate the likelihood of our data $\mathbf{d}$:
\begin{equation}
\label{eq:LikelihoodFull}
\mathcal{L}(\mathbf{d}_{m_a}|A,\mathbf{b}) = \prod_{j = 1}^{\mathcal{N}}\prod_{k = k_i(m_a)}^{k_f(m_a)} P(\bar{\mathcal{F}}_{j,k};N_{{\rm avg},j},\lambda_{j,k}),
\end{equation}
where $k$ indexes the (windowed) frequency bins and $j$ indexes the different spectra. We allow for the generic possibility that each spectrum $\Px_j$ has a different number of averages, $N_{{\rm avg}, j}$. This accounts for the spectra collected at the very end of the data taking period which have a different number of averages.

With the likelihood in Eq.~(\ref{eq:LikelihoodFull}), we perform a likelihood ratio test to search for a possible axion signal. To claim a detection, we place a 5$\sigma$ threshold on the profiled likelihood ratio between the signal-plus-background and background-only hypotheses. We define a \TS for discovery as
\begin{equation}
\label{eq:TSDef}
{\rm TS}(m_a) = 2\ln\left[ \frac{\mathcal{L}(\mathbf{d}_{m_a}|\hat{A},\hat{\mathbf{b}})}{\mathcal{L}(\mathbf{d}_{m_a}|A = 0, \hat{\mathbf{b}}_{A = 0})}\right]\,,
\end{equation}
where $\hat{A}$ and $\hat{\mathbf{b}}$ are the values of $A$ and $\mathbf{b}$ which achieve the global maximum of the likelihood, and $\hat{\mathbf{b}}_{A = 0}$ is the value which achieves the constrained maximization with $A=0$. The maximization of $A$ is performed over a range including positive and negative parameter values, accommodating that a negative parameter value may provide the optimal fit to the data. If $\hat{A} <0$, it is understood that the corresponding best-fit axion coupling is 0, as no value of $g_{a \gamma \gamma}$ could lead to negative-valued $\hat{A}$. 
The $5\sigma$ condition for discovery at a given $m_a$ is ${\rm TS}(m_a) > {\rm TS}_{\rm thresh}$, where
\begin{equation}
\label{eq:LookElsewhere}
{\rm TS}_{\rm thresh} = \left [ \Phi^{-1}\left(1 - \frac{2.87 \times 10^{-7}}{N_{m_a}}\right) \right]^2
\end{equation}
accounts for the local significance as well as the \LEE for the $N_{m_a}$ independent masses in the analysis (here $\Phi$ is the cumulative distribution function for the normal distribution with zero mean and unit variance) \cite{Foster2018}. For this analysis, $N_{m_a} \approx8.1\times10^6$ between 75\,kHz and 2\,MHz (see below), and ${\rm TS}_{\rm thresh} = 56.1$.

Where we do not see a detection, we set a 95\%~C.L.\ limit, $A_{95\%}$, with a similar profiled likelihood ratio.  To do so, we use the following test statistic for upper limits
\begin{equation}
t(m_a,A) = 
\begin{cases}
2 \ln \left[ \frac{\mathcal{L}(\mathbf{d}_{m_a}|\hat{A},\hat{\mathbf{b}})}{\mathcal{L}(\mathbf{d}_{m_a}|A, \hat{\mathbf{b}}_A)}\right] & A\ge\hat{A}\,, \\ 
0 & \mathrm{Otherwise}\,.
\end{cases}
\end{equation}
Here, $\hat{\mathbf{b}}_A$ is the background values that maximizes the likelihood for a given $A$. Using $t$, we can establish the 95\%~C.L.\ limit $A_{95\%}$ where $t(m_a,A_{95\%}) = 2.71$. In this limit setting procedure, it is necessary that $\hat{A}$ was allowed to be negative-valued if this provided the best fit in order to make an accurate calculation of $A_{95\%}$. In addition, we implement one-sided power-constrained limits~\cite{Cowan:2011an}, which in practice means that we do not allow ourselves to set a limit stronger than the $1\sigma$ lower level of the expected sensitivity band. This ensure that our constraints are statistically conservative while also addressing the possibility that $A_{95\%}$ is negative-valued. 

Finally, we discuss the set of mass points over which we scan. In principle, we can search for an axion signal at any value between $75\,\mathrm{kHz}<m_a/(2\pi)<2\,\mathrm{MHz}$. In our data, this range is spanned by $57.5\times10^6$ frequency bins and it might seem natural to search for an axion signal centered at each frequency bin. However, since each axion signal model is resolved by between 10 and 100 frequency bins, neighboring frequency bins would produce very strongly correlated results. Alternatively, a log-spaced set of $N$ masses such that $m_a^{i+1}/m_a^i = 1 + 4 v_0^2$, would achieve a minimal coverage such that every frequency bin belongs to exactly one fit window. However, the spacing that achieves a set of statistically independent axion mass points is given by \mbox{$m_a^{i+1}/m_a^i\approx1+3v_0^2/4$}~\cite{Foster2018}, which yields $N_{m_a}\approx8.1\times10^6$  independent axion masses within our frequency range. For our search procedure, we therefore increase the granularity of the search and produce a set of log-spaced masses that obey $m_a^{i+1}/m_a^i = 1 + v_0^2/2$. This eight-fold enhancement in the resolution of our tested masses, as compared to the minimal coverage set, results in overlapping signal windows of masses studied in our analysis and allows us to over-resolve a potential axion signal by a factor of two. This yields a total of $13.0\times10^6$ mass points to test, which is appropriately larger than the estimated number of independent mass points. We emphasize that this choice of mass points is not a fundamental limit on our mass resolution but is instead imposed merely by computational resources. In the event of an observed excess, we could fit a region around it with the mass floating in the fit.

\begin{figure*}
\centering
\subfloat[\label{fig:VetoData}]{\centering\includegraphics[width=.43\textwidth]{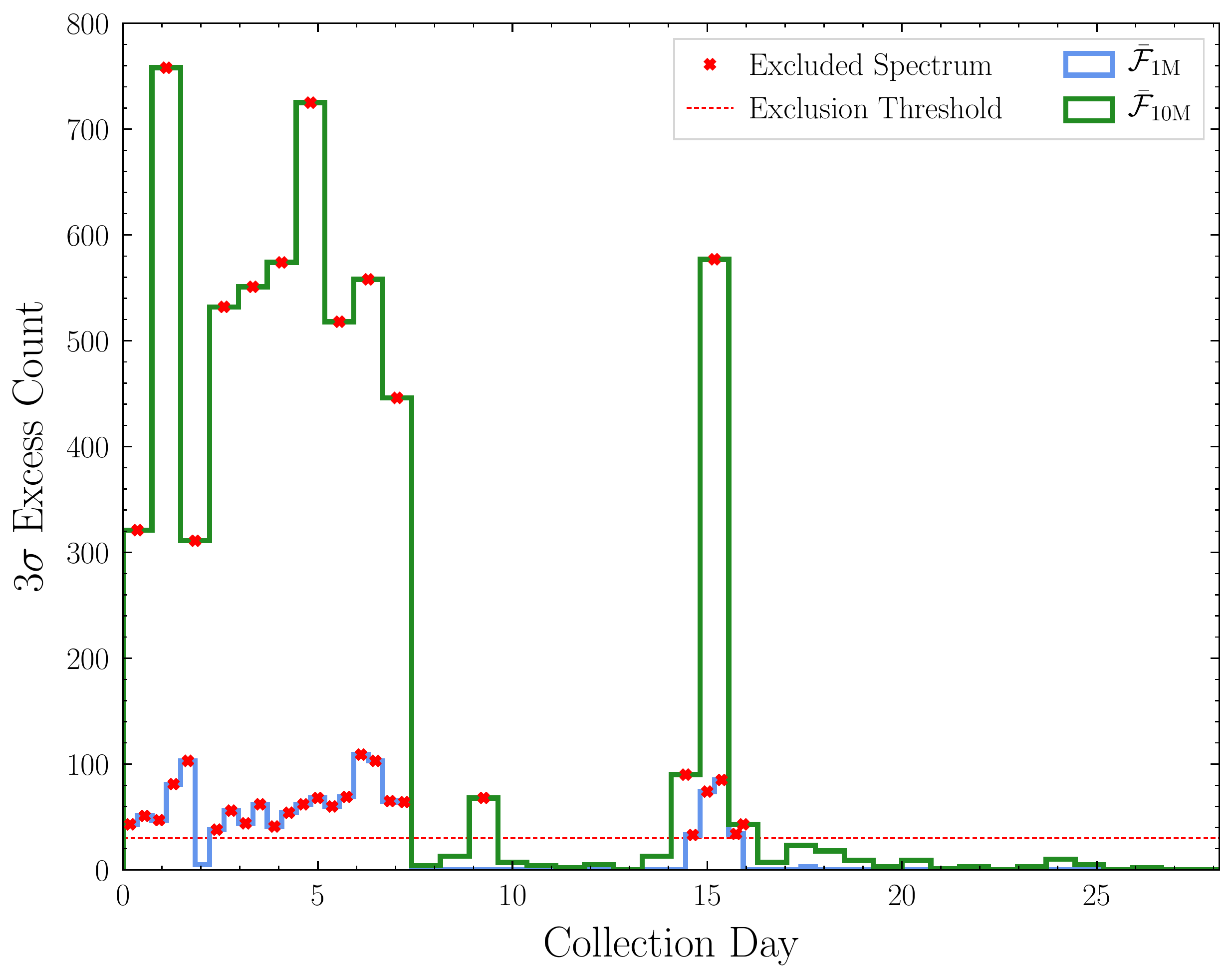}}\hspace{10mm}
\subfloat[\label{fig:SurvivalPlot}]{\centering\includegraphics[width=.43\textwidth]{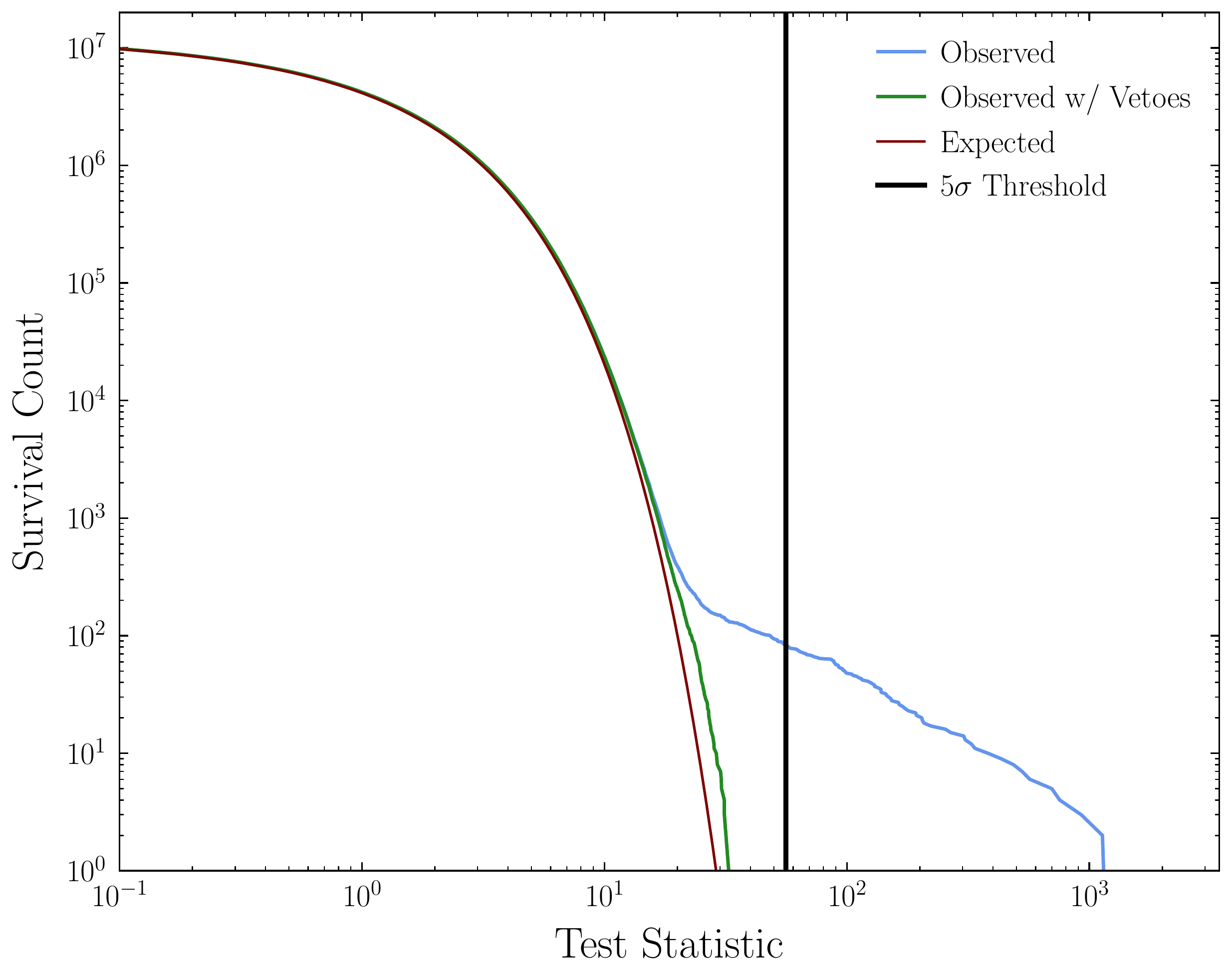}}\caption{(a) The number of 3$\sigma$ excesses accounting for the look-elsewhere effect in each spectrum after vetoing the excesses that are present in the corresponding Magnet Off data. (b) The distribution of local TS values in the full month of analyzed data after removing periods of transient noise. In blue is the observed distribution of local TS values prior to vetoing the Magnet Off excesses. In green, the observed distribution of TS values after the Magnet Off veto. In red, the expected distribution under the null hypothesis. We see that after applying vetoes, there is excellent agreement down to very low survival counts, with no remaining 5$\sigma$ excesses.}
\end{figure*}

For each mass studied, we also compute the expected sensitivity bands from the null-hypothesis models using the Asimov dataset procedure~\cite{Cowan:2010js}, following implementation outlined in~\cite{Foster2018}.

\subsection{Quality Cuts}
\label{sec:QualityCuts}

We can also use the analysis infrastructure described in the previous section to veto mass points where the condition of Gaussian-distributed incoherent noise does not hold as well as to create a quantitative data quality cut to identify periods of time with excess transient noise. 

In order to tag periods of time with increased transient noise, we leverage the fact that the transient noise does not appear as a single peak, but instead as a forest of many correlated peaks. Under the null hypothesis of a flat background, the survival function for the test statistic $t$ is given by
\begin{equation}\label{eq:Survival}
S(t) = 2\left(1 - \Phi\left(\sqrt{t}\right)\right)\,.
\end{equation}
The presence of a true axion signal, would yield a small number of mass points that deviate from this distribution. But a violation of the null hypothesis of a flat background -- e.g.\ due to a forest of correlated transient noise -- would lead to a much larger deviation from this distribution. 

For each $\Px_j$, we calculate the number of mass points with at least a $3\sigma$ excess within the time period covered by that \PSD, accounting for the \LEE. We find that the number of $3\sigma$ excesses, follows a clear bi-modal distribution with an obvious time correlation, see Fig.~\ref{fig:VetoData}. This allows us to place a quantitative cut by requiring that an averaged \PSD, $\Px_j$, have fewer than 30 mass points with a 3$\sigma$ excess. This effectively eliminates periods of time with transient noise. We emphasize two points here: first, by placing this cut on a statistic which is calculated across a broad ranges of frequency, we do not produce a bias at any one mass point or range. Second, as we describe in the next section, a single axion signal would not be expected to create 30 mass points with $3\sigma$ excesses or larger. So, while exotic models with multiple axion could be affected by this cut, it would not present a problem for our baseline model of a single axion. This cut removes $\approx30$\% of our exposure.

Once we have removed periods of time with high transient noise, we remove individual mass points that have non-transient noise peaks -- or are in other ways inconsistent with our null hypothesis of a flat background. We perform our axion discovery analysis on the Magnet Off data -- where we expect no axion signal to be present. Any mass points that show \LEE-corrected excess beyond 5$\sigma$ in the Magnet Off data are vetoed. We consider these mass points to have poorly understood backgrounds where we do not have sensitivity to \ADM. Out of $13.0\times10^6$ mass points, this requirement vetoed 18,733(6,651) points in the range $500\,\mathrm{kHz} < f < 2\,\mathrm{MHz}$ ($75\,\mathrm{kHz} < f < 500\,\mathrm{kHz}$) and implies a decrease in our signal recovery efficiency of 0.2\%.  

The axion search data, collected with the magnet on, showed 83(0) excesses with \LEE corrected significance $\geq5\sigma$, however they were all vetoed by cutting against the Magnet Off data. It is worth pointing out that the number of $5\sigma$ excesses in the axion search data was much smaller than in the Magnet Off data, due to the lower transient noise levels seen during that run. In Fig.~\ref{fig:SurvivalPlot}, we show the distribution of TS values before and after the Magnet Off veto, compared against the distribution expected under the null hypothesis. The strong agreement between the theoretical expectation and the observed distribution after vetoes are applied demonstrates that the experimental backgrounds are well-modeled by the null hypothesis and \abra has strong discovery power under deviations from this theoretical expectation. In particular, in the results presented in \cite{ABRAFirstResults}, we find no significant excesses after vetoes are applied. 

\subsection{Recovering an Injected Signal}
\label{sec:InjectedSignal}

\begin{figure*}
\centering
\includegraphics[width=.9\textwidth]{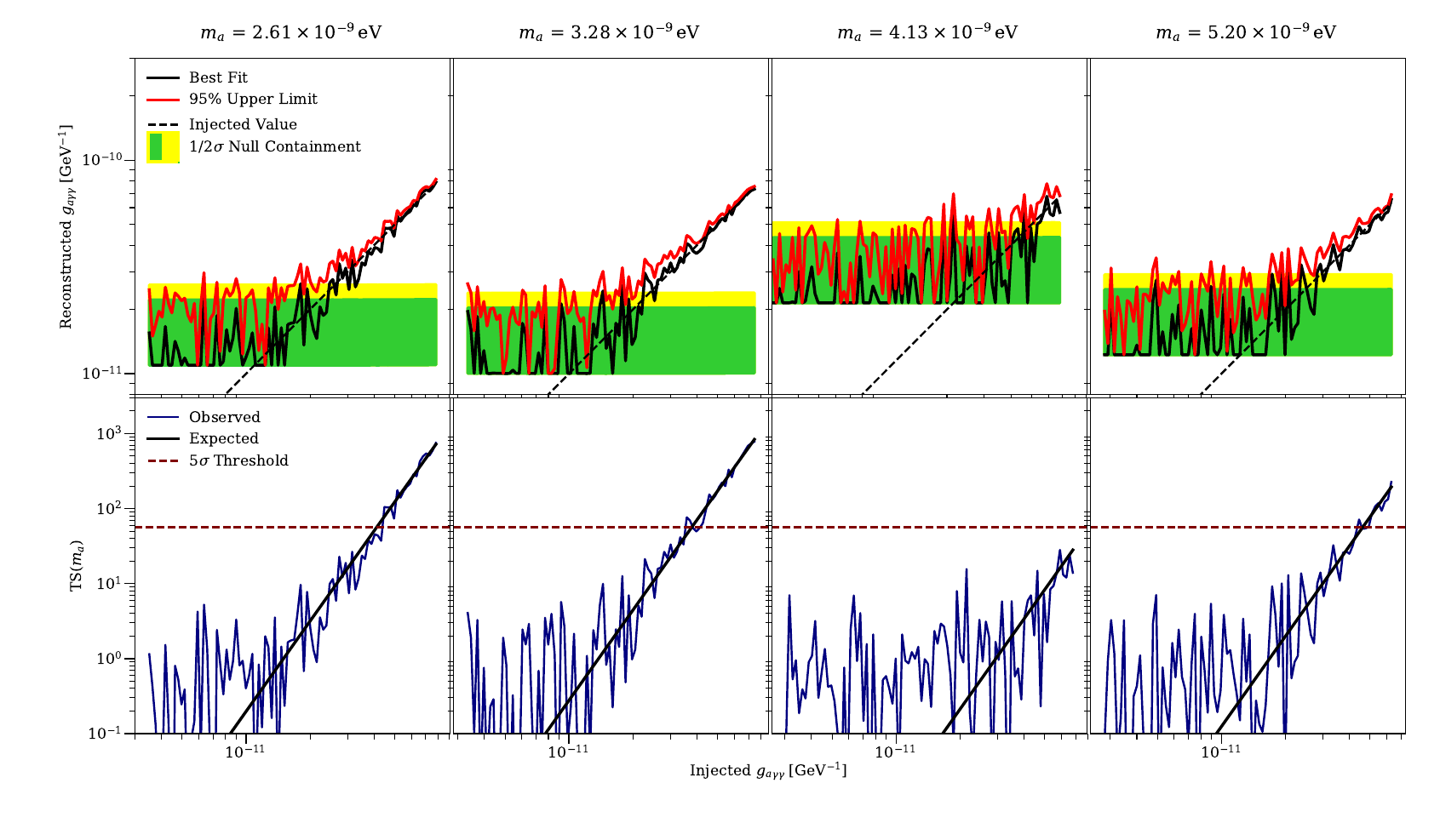}
\caption{(\textit{Top row}) The recovered signal parameters as a function of the injected signal parameters in four Monte Carlo realizations with identical mean background levels. Green and yellow bands indicate the expected 1 and 2$\sigma$ containment for the upper $95\%$ limit on the axion coupling under the hypothesis of no axion signal. (\textit{Bottom row}) The observed and expected test statistic for discovery as a function of the injected signal strength. The dashed red line indicates the threshold for a discovery at $5\sigma$ significance accounting for the LEE, while the dashed black line indicates the upper $95\%$ limit on the observed test statistic under the null hypothesis.}
\label{fig:InjectionTest}
\end{figure*}

As a final test of our analysis procedures, we test that we are able to recover an injected signal and discover an axion at the claimed significance. This is crucial because as axion searches achieve greater sensitivity, there will be an inevitable trade-off between broadband and narrow-band coverage, and a claimed exclusion at a given $\gagg$ will be used as justification to avoid re-scanning parameter space that has already been tested.

To test the discovery power of our analysis procedure, we generate \MC spectra characterized by a mean background level $\hat b$ and \{$N_{{\rm avg}, j}$\} averaged PSDs identical to that of the observed data. We then add an artificial axion signal with signal template set by the SHM, for a range of axion coupling strengths $A$. We perform our quality cuts on the individual spectra, then perform the joint analysis on the surviving spectra. We then evaluate the best-fit axion coupling and the 95$^{th}$ percentile limit on that coupling as a function of the ``true'' axion coupling of the injected signal for each \MC realization. Figure~\ref{fig:InjectionTest} shows the resulting distributions of reconstructed \gagg and $t$-values for six axion masses. 

Critically, the ability of our analysis procedure to accurately recover the correct axion parameters when allowable by the background level is unaffected by the quality cuts. This is most clearly seen in the bottom panels of Fig. \ref{fig:InjectionTest}, which show the $t$-value as a function of the injected signal strength, as compared to the expected $t$-value as a function of the injected signal. We see strong agreement between the expected and observed test statistic when the signal is strong enough that we expect to be able to discover it (i.e., when the injected coupling lies above the null model containment bands). These examples also demonstrate that when our signal is not significant enough to be discovered, our limit-setting procedure is unaffected by the quality cuts.

\section{Conclusions}
The successful run of the \abra experiment \cite{ABRAFirstResults} introduced and validated several new techniques useful for constraining axion dark matter, including a toroidal magnet geometry sensitive to \ADM at $m_a\lesssim1\,\mu$eV, a broadband readout technique capable of handling the data-load required to study millions of axion masses simultaneously, a signal injection through a calibration loop to characterize this type of detector, and the first implementation of the broadband data analysis technique proposed in \cite{Foster2018}. In this paper, we have described in detail the implementation and validation of these techniques, which lend additional to support to the results presented in \cite{ABRAFirstResults}. 

Of greatest practical concern for the first results is the identification of the mismatch between expected and measured end-to-end gain, which we aim to rectify in the next data-taking run, and the mitigation of vibrational noise. We have also emphasized the statistical analysis employed to extract the first results. The goal of this analysis is to establish a sure footing for the presented statistical limits with a robust understanding of the exclusion limits. This is important as next generation experiments reach for ever higher sensitivities and re-scanning regions of parameter space becomes prohibitively time-consuming. In addition, the excellent performance of our data quality cuts will allow use of a blind analysis pipeline, which we expect to use in future runs.

\abra represents the first step in an experimental search program, which aims to ultimately be sensitive to \ADM in the coupling range preferred by QCD axions. Future phases of ABRACADABRA will require larger magnets with higher fields, improved shielding, and strong mitigation of mechanical vibration. Augmenting the techniques described here with a resonant amplification readout and scan strategy will also greatly improve the sensitivity of a future full scale ABRACADABRA detector \cite{ABRA2016,DMRadio_Design,Chaudhuri:2018rqn}. We have already begun engineering studies towards designing and building such a detector and \abra creates a strong foundation for this ongoing work.

\begin{acknowledgments}
\it
The authors would like to acknowledge the useful conversations and advice from Henry Barthelmess of Magnicon Inc. This research was supported by the National Science Foundation under grant numbers NSF-PHY-1658693, NSF-PHY-1806440, NSF-PHY-1505858 and NSF-PHY-1122374. This work was also supported by DOE Early Career Grant number DE-SC0019225, by the Kavli Institute for Cosmological Physics at the University of Chicago, by the Miller Institute for Basic Research in Science at the University of California, Berkeley, by the MIT Undergraduate Research Opportunities in Physics program, by the Leinweber Graduate Fellowship Program and by the Simons Foundation through a Simons Fellowship in Theoretical Physics. This research was supported in part through computational resources and services provided by Advanced Research Computing at the University of Michigan, Ann Arbor. This material is based upon work supported by the U.S. Department of Energy, Office of Science, Office of Nuclear Physics under Award Numbers DE-FG02-97ER41041 and DEFG02-97ER41033 and by the Office of High Energy Physics under grant DE-SC-0012567. We would like to thank the University of North Carolina at Chapel Hill and the Research Computing group for providing computational resources and support that have contributed to these research results.
\end{acknowledgments}

\end{document}